\documentclass[conference]{IEEEtran}
\def\IEEEtitletopspace{18pt}
\IEEEoverridecommandlockouts
\usepackage{cite}
\usepackage{amsmath,amssymb,amsfonts}
\usepackage{textcomp}
\usepackage{amsthm}
\usepackage{bbold}
\usepackage{array}
\usepackage{hyperref}
\usepackage{caption}
\usepackage{soul}
\usepackage{color}
\usepackage[pdftex]{graphicx}
\usepackage{subcaption}
\usepackage{enumitem}
\usepackage{cite}
\usepackage{algorithm}
\usepackage{algpseudocode}
\algrenewcommand\algorithmicrequire{\textbf{Input:}}
\algrenewcommand\algorithmicensure{\textbf{Output:}}
\usepackage{subfiles}
\usepackage{verbatim}
\usepackage{bbm}
\newtheorem{theorem}{Theorem}

\newtheorem{proposition}{Proposition}
\newtheorem{lemma}{Lemma}
\usepackage[font = small, justification = justified,format = plain]{caption}

\usepackage{xcolor}
\def\BibTeX{{\rm B\kern-.05em{\sc i\kern-.025em b}\kern-.08em
    T\kern-.1667em\lower.7ex\hbox{E}\kern-.125emX}}
\begin{document}

\title{On the Optimality of Procrastination Policy for EV Charging under Net Energy Metering}

\author{Minjae~Jeon,~\IEEEmembership{Student Member,~IEEE,}
        Lang~Tong,~\IEEEmembership{Fellow,~IEEE,}
        Qing~Zhao,~\IEEEmembership{Fellow,~IEEE}
        
    \thanks{Minjae Jeon, Lang Tong, and Qing Zhao(\{\textcolor{blue}{\texttt{mj444, lt35, qz16}}\}\textcolor{blue}{\texttt{@cornell.edu}}) are with the School of Electrical and Computer Engineering, Cornell University, USA. This work was supported in part by the National Science Foundation under Grant 2218110 and 1932501.}
    }

\maketitle
\begin{abstract}
    We address the problem of behind-the-meter EV charging by a prosumer, co-optimized with rooftop solar, electric battery, and flexible consumption devices such as water heaters and HVAC systems. We consider a scenario involving a time-of-use net energy metering tariff, alongside stochastic solar production and random EV charging demand, a finite-horizon surplus-maximization problem is formulated. We show that a procrastination threshold policy that delays EV charging to the last possible moment is optimal when EV charging is co-optimized with flexible demand, and the policy thresholds can be easily computed offline. When battery storage is part of the co-optimization, we show that the prosumer's net consumption is a two-threshold piecewise linear function of the behind-the-meter renewable generation under the optimal policy, and the procrastination threshold policy remains optimal, although the thresholds cannot be computed easily. We propose a straightforward myopic solution which becomes optimal strategy when the storage state of charge doesn't reach boundary, and demonstrate in a numerical simulation that involves real-world data.
\end{abstract}
\begin{IEEEkeywords}
EV charging, distributed energy resources, stochastic dynamic programming, net energy metering 
\end{IEEEkeywords}
\section{Introduction}
    We consider the problem of co-optimizing electric vehicle (EV) charging, flexible consumption devices, behind the meter (BTM) rooftop solar, and energy storage. This work is motivated by the growing adoption of the BTM DER and storage \cite{solarstoragereport} to shift and flatten the aggregated household energy consumption. With EV as a deferrable load, co-optimizing BTM resources benefits not only individual prosumers but also the distribution grid operations in reduced power flow \cite{alahmed2022net}. 

    Most households with BTM DER in the U.S. are under some form of the net energy metering (NEM) tariff offered by a regulated utility or consumer choice aggregator, which bills the prosumer for its net energy consumption reading from the revenue meter, as shown in Fig. 1. We assume that an energy management system (EMS) controls flexible household demands, such as water heater and HVAC, EV charging, and BTM storage based on available renewable generation. 

    The co-optimization problem can be formulated with continuous state and action spaces Markov decision process (MDP). Unfortunately, the solution to such a stochastic dynamic program (DP) is intractable without exploiting the special properties of the problem. To this end, we focus on the particular structure of the NEM 2.0 and beyond \cite{alahmed2022net}, where the purchasing (importing) rate is higher than the selling (exporting) rate. A striking property of such NEM tariffs is that it creates a net-zero zone in the household's net consumption. Our work in this paper builds upon these structural properties uncovered recently in \cite{alahmed2022co,alahmed2022net}.

    \begin{figure}[t]
    \centerline{\includegraphics[width = \linewidth]{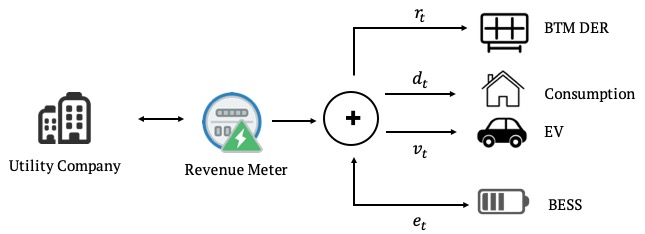}}
    \caption{NEM scheme of the household with the BTM storage and EV. The direction of arrow indicates the direction of the power flow.}
    \label{fig1}
    \vspace{-1em}
    \end{figure}

\subsection{Related works}
    This work is closely related to \cite{jeon2022co, alahmed2022co} which studied the joint scheduling of EV and flexible loads \cite{jeon2022co}, and the joint scheduling of energy storage and flexible loads \cite{alahmed2022co}. A considerable amount of research has also been conducted on the joint scheduling of BTM battery with an EV \cite{hafiz2019coordinated, Seal2023, yousefi2020predictive, chen2013mpc, xu2020multi, Pedrasa2010, Erdinc2015} and flexible demands\cite{Hubert2012, Li2018, Xu2017, Yu2020}, along with DER. However, only a few studies have specifically incorporated the EV charging deadline and the deferrable nature of EV into the co-optimization with storage. Moreover, to the best of our knowledge, no previous literature has explored the online scheduling of EV, storage, and consumption under NEM tariffs. In this article, we examine previous works that have investigated comparable optimization frameworks and techniques.

    The co-optimization problem for the EMS with uncertainties (e.g. rooftop solar generation and outside temperature) was modeled as an one-shot optimization problem in \cite{Pedrasa2010, Hubert2012, Erdinc2015}. However, this approach heavily depends on the accuracy of the forecasts and does not provide an online policy. The authors of \cite{Pedrasa2010} aimed to maximize household surplus by co-optimizing flexible loads with BTM DER. In \cite{Erdinc2015, Hubert2012}, the authors formulated an mixed integer linear programming (MILP) for the co-optimization but didn't consider controlling flexible loads, and EV charging, respectively. 

    Several methods formulated the co-optimization problem as an MDP\cite{hafiz2019coordinated, Kim2011, Xu2017, alahmed2022co, jeon2022co}. Most of these works did not provide a policy simultaneously scheduling EV charging, BTM storage, renewables, and flexible demands. To avoid `curse of dimensionality', authors of \cite{Kim2011, Xu2017, alahmed2022co, jeon2022co} exploited threshold structure of the optimal policy. However, in \cite{alahmed2022co ,Xu2017}, EV was not part of the optimization, and in \cite{jeon2022co} storage wasn't modeled. Also, in \cite{Kim2011}, the preference of the different flexible loads were not considered. 

    Model predictive control (MPC) has been widely studied as an alternate approach for solving MDP for joint scheduling under uncertainties \cite{chen2013mpc, Seal2023, yousefi2020predictive}. MPC can be adapted in real-time using the current available information but it's performance highly depends on the forecasting algorithm and problem size, such as look ahead horizon and state/action space. 

    Recently, reinforcement learning (RL)-based methods have been applied for home EMS co-optimization in \cite{wu2018optimizing, Yu2020, xu2020multi}. The RL-based approach has the advantage of not requiring knowledge about uncertainties, but it is not free from the `curse of dimensionality' for high-dimensional discrete or continuous state/action space. For instance, \cite{xu2020multi, wu2018optimizing} Q-learning-based method was applied to the discretized state-action space co-optimization problem which can not be scaled to large state-action space. In \cite{Yu2020}, the authors used deep deterministic policy gradient (DDPG) algorithm for the continuous state-action space joint scheduling problem, but the algorithm converges slowly, which restricts its usage in an online setting.

\subsection{Summary of results and contributions}
    The main contribution of this work is threefold. First, we show that the \textit{procrastination threshold policy}, that delays the EV charging until the last possible moment is optimal for the co-optimized scheduling of EV charging, flexible load, and stochastic renewable generation. We then propose a myopic solution for the BTM battery operations and show that the decisions under the myopic policy become optimal when the storage state of charge (SoC) doesn't reach its boundary.

    Second, we show that the myopic storage operations are piecewise linear function of renewable generation which is divided into 5 segments by 4 thresholds. The myopic battery operations store energy in the battery solely using BTM DER generation, and discharge only to curtail the net consumption of the household.
        
    Finally, we provide results from an empirical study using the real-world data to show that the procrastination threshold policy with myopic storage operations outperforms MPC and other non co-optimization policies and are within 0.5-5\% performance gap to an oracle policy.

The notations used in the paper are standard. Vectors are in boldface, with $(x_1, \ldots, x_N)$ a column vector. We use $\mathbf 1$ for a column of ones with appropriate size. $\mathbb 1_A$ is an indicator function that maps to 1 if A is true, and zero otherwise. 

The proofs of theorems and propositions are omitted due to space limitations, and they can be found at \cite{jeon2023Procras}.

\section{Problem Formulation}
We consider a sequential scheduling of EV charging $(v_t)$, flexible household devices $(\mathbf d_t)$, and storage operation $(e_t)$ over a discrete-time and finite horizon of length $T$ with intervals indexed by $\mathcal T = \{0, \ldots, T-1\}$.

\subsection{EV charging, consumption, and storage model}

\paragraph{BTM DER $(r_t)$}  We model the BTM DER generation during interval $t$ as an exogenous variable, represented by a sequence of independent random variables with a time-dependent distribution $f_t(\cdot)$. We assume that the realization of DER is known at the beginning of each scheduling interval.
\begin{equation}
    r_t \sim f_t(\cdot), \quad \forall t \in \mathcal T
\end{equation}

\paragraph{Remaining charging demand $(y_t)$}  At the start of every interval, the remaining charging demand, $y_t$, is measured. A charger with a maximum capacity $\bar v$ and charging efficiency $\eta$ provides $v_t$ in interval $t$. We do not allow discharging the EV battery and the total energy supplied to the EV will not exceed the amount requested at the beginning.
\begin{align}
    y_{t+1} &= y_t - \eta v_t,\quad \forall t \in \mathcal T.\\
    y_t &\in [0, y_0], \quad \forall t \in \mathcal T. \\
    v_t &\in [0, \min \{y_t / \eta, \, \bar v\}], \quad \forall t \in \mathcal T.
\end{align}
Without loss of generality we assume $\eta = 1$. \footnote{Rescale $y_0$ and $y_t$ by $1/\eta$.}

\paragraph{Household consumption $(\mathbf d_t)$}  Consumption vector $\mathbf d_t=\big(d_{t1},\cdots,d_{tI}\big)$ models the flexible consumption of $I$ controllable devices within interval $t$ which satisfies : 
\begin{align}
    0 \preceq \mathbf d_t \preceq \mathbf{\bar d} := (\bar d_{1}, \ldots, \bar d_I), \quad \forall t \in \mathcal T.
\end{align}

We model the utility of consuming $\mathbf d_t$, $U_t(\mathbf d_t)$, as an additive concave function with a marginal utility $L_t(\mathbf d_t)$ : 
\begin{equation*}
    U_t(\mathbf d_t) = \sum_{i=1}^I U_{ti}(d_{ti}), \; L_t(\mathbf d_t)= \big(L_{t1}, \ldots, L_{tI}\big) := \nabla U_t  .
\end{equation*}

Uncontrollable loads are not explicitly modeled here; they are known and subtracted from the renewable generation.

\paragraph{Battery storage operation $(e_t)$} The battery SoC is denoted as $s_t \in [0, B]$ where $B$ represents the capacity of the storage. The battery operation is denoted as $e_t \in [-\underline{e}, \bar e]$, where $e_t \ge 0$ represents charging, and $e_t \le 0$ represents discharging. Battery charging and discharging efficiency constants are denoted as $\eta_c, \eta_d \in (0, 1]$, respectively. 
\begin{align}
    e_t &\in [-\underline e, \bar e], \quad \forall t \in \mathcal T\\
    s_t &\in [0, B] \label{soclimit},  \quad \forall t \in \mathcal T\\
    s_{t+1} &= s_t + (\mathbb 1_{e_t \ge 0}\eta_c  + \mathbb 1_{e_t \le 0} / \eta_d)e_t, \quad \forall t \in \mathcal T.
\end{align}

\paragraph{Net consumption ($z_t$)}  The net energy consumption of a household within interval $t$ is defined as the net energy consumption measured by the revenue meter during the billing period. This includes EV charging, flexible loads, storage operation and BTM DER generation within the billing period.
\begin{equation*}
    z_t := v_t + \mathbf 1 ^T\mathbf d_t + e_t - r_t, \quad t \in \mathcal T.
\end{equation*}
We say that household is \textit{net-consuming} when $z_t > 0$, and \textit{net-producing} when $z_t < 0$.

\subsection{NEM ToU tariff model} 
\paragraph{NEM payment $\big(P^{\pi_t}(\cdot)\big)$} Under the NEM tariff program, a household  is billed or credited based on the net energy consumption during each billing period, which can range from five minutes to a day or a month. To simplify the notation, we matched the length of our decision intervals with the billing period that computes the net consumptions.

At interval $t$, a household with net consumption $z_t$ pays 
\begin{equation*}
    P^{\pi_t}(z_t) := z_t(\mathbb 1_{z_t \ge 0} \pi_t^+ + \mathbb 1_{z_t \le 0} \pi_t^-)+ \pi^0_t,
\end{equation*}
where $\pi_t^+$ is a \textit{retail rate},  $\pi_t^-$ is a \textit{sell rate}, and $\pi_t^0$ is a fixed charge\footnote{Fixed charge doesn't affect the optimal decision, so we assume $\pi_t^0 = 0$.}. A prosumer pays the retail rate for the net consumption and credited at the sell rate for the net generation. See \cite{alahmed2022net}. 

\begin{figure}[t]
\centerline{\includegraphics[width = 0.9\linewidth]{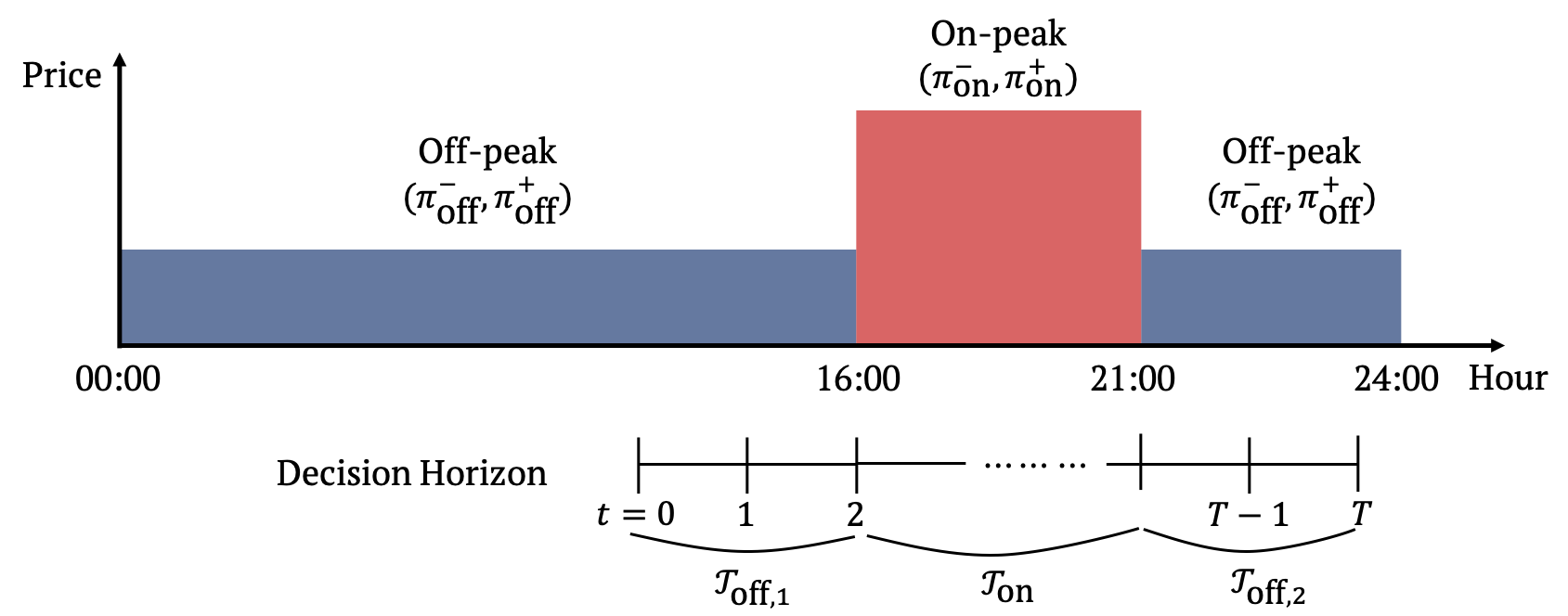}}
\caption{ToU scheme and decision horizon}
\label{fig2}
\vspace{-1em}
\end{figure}

\paragraph{NEM ToU tariff model} ToU tariff divides 24 hours into periods with different energy prices. For our analysis, we adopt a ToU tariff with two distinct energy prices : off-peak and on-peak\footnote{A typical ToU tariff sets the on-peak hours to a five-hour period in the late afternoon and early evening as in Fig.\ref{fig2}}. We assume that the decision horizon falls within the first off-peak period, the on-peak period, and the second off-peak period as shown in Fig.\ref{fig2}\footnote{We refer period as the set of consecutive intervals with the same price.}. Hence, once we know the EV connection time and the deadline, the scheduling horizon is well defined : $\mathcal T = \mathcal T_{\text{off},1} \cup \mathcal T_{\text{on}} \cup \mathcal T_{\text{off},2}$, with $T = |\mathcal T_{\text{off},1}| + |\mathcal T_{\text{on}}| + |\mathcal T_{\text{off},2}|$.

We assume the typical implementation where the prices are the same within the on- and off-peak periods. Standard tariff designs avoid risk-free arbitrage, which requires that NEM parameters satisfy $\pi_{\text{off}}^-  <\pi_{\text{on}}^-< \pi_{\text{off}}^+ < \pi_{\text{on}}^+$.
\begin{equation}
    \pi_t = \begin{cases}
        \pi_{\text{on}} := ( \pi_{\text{on}}^-, \pi_{\text{on}}^+),
        & t \in \mathcal T_{\text{on}}.\\
        \pi_{\text{off}} := ( \pi_{\text{off}}^-, \pi_{\text{off}}^+), 
        & t \in \mathcal T_{\text{off,1}} \cup \mathcal T_{\text{off,2}}. \\
    \end{cases}
\end{equation}

\subsection{Co-optimization of EV charging, consumption, and storage}
    We formulate the scheduling problem as a stochastic DP with state $x_t = (s_t, y_t, r_t, \pi_t)$ and action $a_t = (v_t, e_t, \mathbf d_t)$. The initial SoC $(s_0)$ and EV charging demand $(y_0)$ are given. 
    
    The policy $\mu := (\mu_0, \ldots, \mu_{T-1})$ is a sequence of functions that maps the current state to action $a_t$ :  
    \begin{equation}
        \mu_t(x_t) := (v_t, e_t, \mathbf d_t), \quad  t \in \mathcal T.
    \end{equation}

    The stage reward for the stochastic DP is the household surplus under the NEM tariff with the terminal reward being the sum of the salvage value of battery storage and the penalty for incomplete EV charging demand. The salvage value of battery storage is proportional to the terminal SoC, and the penalty for incomplete EV charging is proportional to the remaining charging demand at $t = T$.
\begin{equation*}
    g_t(x_t, v_t, e_t, \mathbf d_t) := 
    \begin{cases}U_t(\mathbf d_t) - P^{\pi_t} (z_t), & t = 0, \ldots, T-1. \\
    \beta s_T -\alpha y_T, & t = T.
    \end{cases}
\end{equation*}
The co-optimization scheduling problem is given by :
\begin{gather}\label{co_opt_sdp}
    \begin{aligned}
    \mathcal P : & \max_\mu &&\mathbb E_\mathbf r\left[\sum_{t=0}^{T-1} g_t(x_t, v_t, e_t, \mathbf d_t) - \alpha y_T + \beta s_T \right] \\
    & \text{s.t.} && (1)-(10).
    \end{aligned}
\end{gather}

We denote the optimal value function as $V_t(x_t)$ that satisfies following Bellman Equation :  
\begin{gather}\label{eq:dp}
\begin{aligned}
     V_t(x_t) &= \max_{v,e,\mathbf d} \;\{g_t(x_t, v, e, \mathbf d) + \mathbb E[V_{t+1}(x_{t+1})] \}  \\
     & \text{s.t. }  (1)-(9)
\end{aligned}
\end{gather}
    We assume that the NEM tariff parameters, the incomplete charging demand penalty, and the battery salvage value satisfy
    \begin{equation} \label{eq:price_penalty_assumption}
    \pi_{\text{off}}^- < \pi_{\text{on}}^- < \eta_c \beta  < \beta / \eta_d< \pi_{\text{off}}^+ < \pi_{\text{on}}^+ < \alpha. 
    \end{equation}

    We assume high penalty to minimize incomplete charging demand. The penalty can be interpreted as the price of fulfilling incomplete charging demand at the deadline. The storage salvage value assumption is to avoid the trivial storage operation where the storage is always charging or discharging.

\section{Procrastination threshold policy (co-optimization without battery)}\label{procras_pol}
We first introduce the \textit{procrastination threshold policy} which is the optimal EV owner’s decision without a BTM battery to provide an intuition for the more general problem.

\begin{theorem}[Procrastination threshold policy] The optimal EV charging $v_t^*$ and consumption $d_{ti}^*$ decisions are monotone increasing function of $r_t$, and for all $i$ and $t$
\begin{align*}
    v_t^* &= 
    \begin{cases}
        h_{\tau_t}(y_t), & r_t < \Delta_t^+(y_t) \\
        h_{w_{t+1}(\nu)}(y_t), & \Delta_t^+(y_t) \le r_t \le \Delta_t^-(y_t) \\
        h_{\delta_t}(y_t), & r_t > \Delta_t^-(y_t).
    \end{cases}\\
    d_{ti}^* &= 
    \begin{cases}
        l_{ti}(\pi_t^+), & r_t < \Delta_t^+(y_t) \\
        l_{ti}(\nu), & \Delta_t^+(y_t) \le r_t \le \Delta_t^-(y_t) \\
        l_{ti}(\pi_t^-), & r_t > \Delta_t^-(y_t),
    \end{cases} 
\end{align*}
where $\nu \in [\pi_t^-, \pi_t^+]$ satisfies, $r_t = v_t^* + \sum_{i=1}^I d_{ti}^*$.\\
$\tau_t$ and $\delta_t$ are characterized by : 
    \begin{align*}
        \tau_t &= 
        \begin{cases}
        (T-t-1) \bar v, & t \in \mathcal T_{\mathrm{off,2}} \cup \mathcal T_{\mathrm{on}} \\
        w_{t+1}(\pi_{\mathrm{off}}^+), & t \in \mathcal T_{\mathrm{off,1}}
        \end{cases}\\
        \delta_t &= \begin{cases}
             \raisebox{-1.6ex}{$0$,} & t \in \mathcal T_{\mathrm{off,1}} \cup \mathcal T_{\mathrm{off,2}},\\
            & t \in \mathcal T_{\mathrm{on}} \land \mathcal T_{\mathrm{off,2}} = \emptyset \\
            w_{t+1}(\pi_{\mathrm{on}}^-), & t \in \mathcal T_{\mathrm{on}} \land \mathcal T_{\mathrm{off,2}} \neq \emptyset
        \end{cases}
    \end{align*}
\begin{flalign*}
    &&\bar V_t(y) &:= \mathbb E[ \tilde V_t(y, r_t, \pi_t)], \quad w_t(\pi) :=(\partial \bar V_t)^{-1}(-\pi), &\\
    && h_{\theta}(y) &:= \min \{ \bar v, \max \{ y - \theta, 0\}\}, & \\
    &&l_{ti}(\pi) &:= \min \{ L_{ti}^{-1}(\pi), \bar d_i\}, \quad l_t(\pi) := \sum_{i =1}^I l_{ti}(\pi), &\\
    && \Delta_t^-(y_t) &:=l_t(\pi_t^-) + h_{\delta_t}(y_t)\\
    && \Delta_t^+(y_t) &:= l_{t}(\pi_t^+) + h_{\tau_t}(y_t).& \qedsymbol
\end{flalign*}
\end{theorem}   
    The optimal value function for co-optimization problem without storage is represented as $\tilde V_t$ and $\partial \bar V_t$ represents subdifferential of the $\bar V_t$.

    The policy is governed by two thresholds on remaining charging demand, $\tau_t$ and $\delta_t$. These thresholds indicate the point at which EV is charged using grid purchase and DER, respectively. Charging level for each zone is dictated by the difference between $y_t$ and the thresholds if $y_t$ surpasses them. 
    
\begin{figure}[h]
    \vspace{-1em}
    \centerline{\includegraphics[width = 0.9\linewidth]{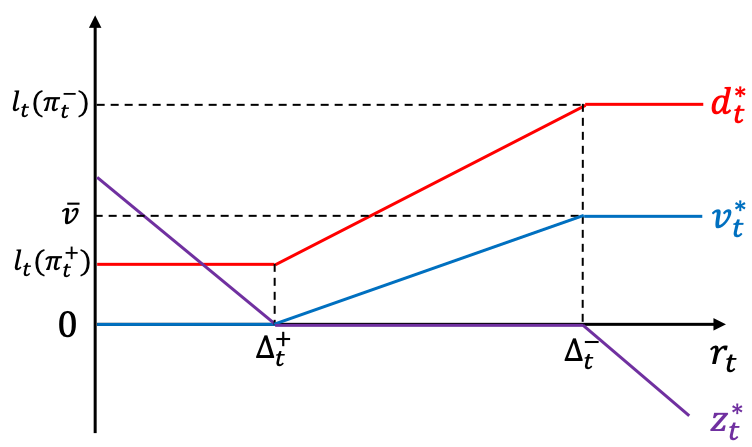}}
    \caption{Procrastination threshold policy for $\delta_t < y_t < \tau_t$.}
    \vspace{-0.5em}
    \label{fig3}
    \end{figure}

    The combination of EV charging level and surplus maximizing consumption levels for net-consuming and -producing zone give thresholds for BTM DER, $\Delta_t^+(y_t)$ and $\Delta_t^-(y_t)$, which decide household's net consumption state (net-consuming, net-producing or net-zero). The EV charging and total consumption ($d_t^* = \mathbf 1 ^T \mathbf d_t^*$) decisions with respect to $r_t$ is depicted in Fig.\ref{fig3}. So for $\delta_t < y_t < \tau_t$, EV is not charged by purchasing energy $(z_t^* > 0)$ and charged only if there exists excess DER ($r_t > \Delta_t^+$), as depicted in Fig.\ref{fig3}.

    The naming of the procrastination threshold comes from $\tau_t = (T-t-1)\bar v$ which is equivalent to maximum charging capacity of remaining intervals. Such procrastination threshold is optimal when it's indifferent to purchase energy in current or subsequent intervals. The intuitive explanation for the optimality of procrastination charging behavior is that procrastinating charging behavior increases the likelihood of finishing the job using DER, which is a cheaper charging option. 

    Following proposition shows procrastination behavior of purchasing energy to charge EV. 
    \begin{proposition}[Procrastination charging behavior]
        Within the same pricing period, $\tau_t$ and $\delta_t$ satisfy 
        \begin{equation*}
            \tau_t = \tau_{t+1} + \bar v,\,
            \delta_t = \delta_{t+1},
        \end{equation*}
     and the optimal charging decisions satisfy 
    	\begin{center}
    	    If $z_{t_0}^* v_{t_0}^* > 0$ for $t_0 \in \mathcal T_x$  then $v_t^* = \bar v,\; t_0 < \forall t \in \mathcal T_{x}$ ,  
    	\end{center} 
     where $\mathcal T_x$ is $\mathcal T_{\mathrm{off,1}}, \mathcal T_{\mathrm{off,2}}, \mathcal T_{\mathrm{on}}$. \hfill \qedsymbol
	\end{proposition}
    \vspace{-0.5em}

\section{Optimal prosumer decisions}
We now present a prosumer's decision for EV charging, flexible loads, BTM renewable and storage. We will use the result from the previous section and consider a co-optimization with myopic storage operations to derive the simple storage operation policy. Let's consider following co-optimization problem with myopic storage operation : 
\begin{equation} \label{eq:myopic_opt}
     \begin{aligned}
    V^M_t(x_t) = &\max_{v,e,\mathbf d}  \quad  g_t(x_t, v, e, \mathbf d) +\big(\eta_c \mathbb 1_{e \ge 0} + \mathbb 1_{e\le 0} / \eta_d\big)\beta e \\
     & \quad \quad \quad  +  \mathbb E[V^M_{t+1}(s_t,y_t - v, r_{t+1}, \pi_{t+1})]  \\
     &\text{s.t.} \quad -\min \{\underline e, s_t \eta_d\} \le e  \le \min\{ \bar e, (B - s_t) / \eta_c\}, \\
     &  \quad \quad \; (1) - (5), (9).
    \end{aligned}
\end{equation}
where $V^M_{t}(x_t)$ is the optimal value function for the myopic co-optimization. The second term in the objective function captures the terminal value of the storage operation. The storage SoC constraints are taken into account by clipping storage charging/discharging levels by SoC limits.

\begin{theorem}[Myopic policy] The optimal net consumption $z_t^*$ of myopic co-optimization is a piecewise linear monotone decreasing function of $r_t$ : 
\begin{equation*}
    z_t^* = \begin{cases}
        \Delta_t^{+'}(y_t) - r_t, & r_t < \Delta_t^{+'}(y_t) \\
        0, & \Delta_t^{+'}(y_t) \le r_t \le \Delta_t^{-'}(y_t) \\
        \Delta_t^{-'}(y_t) - r_t, & \Delta_t^{-'}(y_t) < r_t.
    \end{cases}
\end{equation*}
The optimal EV charging $v_t^*$, total consumption $d_t^*$ and the storage $e_t^*$ decisions are monotone increasing functions of $r_t$, segmented by 6 thresholds on $r_t$ : $\Delta_t^{+'}(y_t), \Delta_{t,1}(y_t)-\Delta_{t,4}(y_t), $ and $\Delta_t^{-'}(y_t)$, and they are decided by \textup{\textbf{Algorithm 1}}.
\begin{align*}
    \Delta_t^{+'}(y_t) &:= \max \{ \Delta_t^+(y_t) - \underline e', 0\}, \\
    \Delta_{t,1}(y_t) &:= \max \{ l_t(\beta / \eta_d) + h_{\sigma_t^+}(y_t) - \underline e', 0\}, \\
    \Delta_{t,2}(y_t) &:= l_t(\beta / \eta_d) + h_{\sigma_t^+}(y_t), \\
    \Delta_{t,3}(y_t) &:= l_t(\beta \eta_c)+ h_{\sigma_t^-}(y_t), \\
    \Delta_{t,4}(y_t) &:= l_t(\beta \eta_c) + h_{\sigma_t^-}(y_t) + \bar e', \\
    \Delta_t^{-'} (y_t) &:= \Delta_t^-(y_t) + \bar e',
\end{align*}
and $\sigma_t^+ = w_{t+1}(\beta / \eta_d)$ and $\sigma_t^- = w_{t+1}(\beta \eta_c)$, where $\tau_t > \sigma_t^+ > \sigma_t^- > \delta_t$ for all $t \in \mathcal T$.
\hfill \qedsymbol
\end{theorem}
    Due to page limit, \textbf{Algorithm 1} is in the appendix of \cite{jeon2023Procras}.

    The EV charging and consumption decisions for the net consuming and producing zones are consistent with the optimal policy in Sec.\ref{procras_pol}; only the thresholds on DER which determine net consumption state ($\Delta_t^{+'}$ and $\Delta_t^{-'}$) are adjusted by maximum storage charging and discharging level. Storage operations in these zones are maximum discharging and charging level, respectively. 

    The two new thresholds for remaining charging demand are introduced here : $\sigma_t^+$ and $\sigma_t^-$. These thresholds determine the priority rule between EV charging and storage discharging and charging in the net-zero zone, respectively.

\begin{figure}[h]
        \vspace{-1em}
        \centering
        \includegraphics[width = 0.9\linewidth]{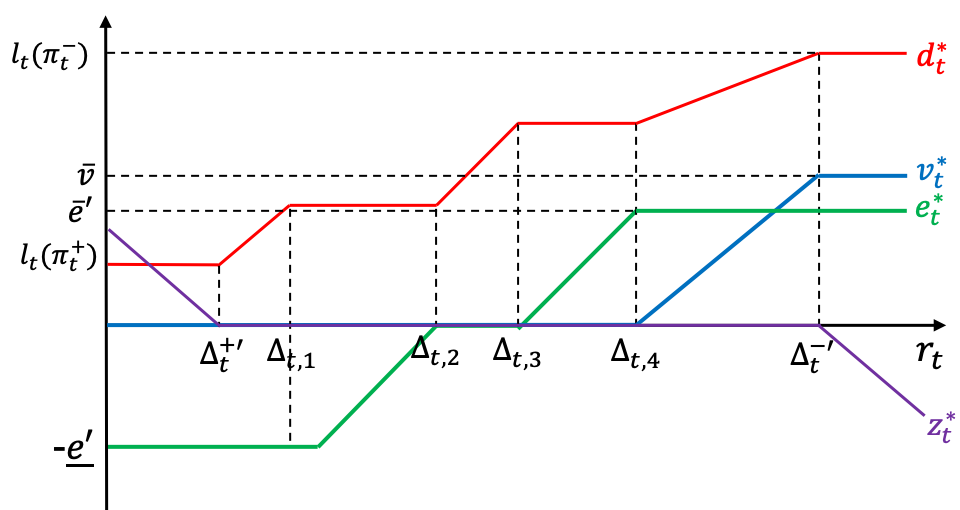}
        \captionof{figure}{Myopic optimal policy for $\delta_t = 0 < y_t < \sigma_t^- < \sigma_t^+$.}
        \label{fig4}
        \vspace{-0.5em}
    \end{figure}

    When $y_t < \sigma_t^+ $, EV charging has lower priority to discharging the storage. This means that the EV is not charging when the storage is discharging. This might happen when the remaining EV charging demand is small or deadline is still far away. For instance, in the Fig.~\ref{fig4}, $v_t^* = 0$ when $e_t^* < 0$. Similarly, for $y_t < \sigma_t^-$, the priority is given to charging the storage over charging the EV, which means that the EV is charged only when the storage is charged at the maximum rate as depicted in Fig.~\ref{fig4} ($r_t > \Delta_{t,4})$.

    When the DER is in between $\Delta_t^{+'}$ and $\Delta_t^{-'}$, household is in the net-zero zone. In this region, the total household load matches $r_t - e_t^*$, BTM DER adjusted by storage output. For instance, if $r_t \in \big[\Delta_t^{+'}(y_t), \Delta_{t,1}(y_t)\big)$, $v_t^* + d_t^* = r_t + \underline e'$. The BTM DER adjusted by storage output is allocated to each device so that marginal utility and marginal value of EV charging is matched.

    The optimal myopic storage operation is a piecewise linear function and it satisfies : $e_t^* z_t^* \le 0.$ This implies that the energy is stored in the battery only using the BTM DER, and only discharged to reduce the net consumption. 

    We now present the optimality of the myopic policy.

    \begin{theorem}[Optimality of myopic policy]
    The action $a_t = (v_t, e_t, \mathbf d_t)$ under myopic policy in \textbf{\textup{Theorem 2}} is the optimal decision of \textup{(\ref{co_opt_sdp})}, if the battery SoC limit constraint, \textup{(\ref{soclimit})}, is nonbinding.
    \end{theorem}

    One of the sufficient conditions that the battery SoC doesn't reach the boundary is $s_0 \in (\underline e T/\eta_d , B - \eta_c \bar e T)$ or $B > T(\eta_c \bar e + \underline e / \eta_d)$. This assumption is valid if $B \gg \eta_c \bar e$ and $ B \gg \underline e/\eta_d $ which corresponds to the storage capacity being large relative to the maximum storage charging and discharging rate.

\section{Numerical results}

    We implemented a numerical simulation using real-world data, which involves a household with the BTM storage and EV under the utility's NEM tariff, to verify the performance of the myopic optimal policy presented in this paper and demonstrate the benefits of the co-optimization.

    We compared the performance of different control methods using the relative expected accumulated surplus gap to the oracle policy. The oracle policy is the offline policy which is the outcome of solving convex optimization (\ref{co_opt_sdp}) with the realizations of renewables, which will serve as the upper bound. If the oracle policy, $\mu_o$ and comparing policy $\mu$ has expected accumulated reward of $R_{\mu_o}$, and $R_\mu$, respectively, the performance gap (in percentage) is given by $(R_{\mu_o} - R_\mu) / R_{\mu_o}$

    The myopic optimal policy (MO) in Theorem 2 is compared with an MPC-based control method and three alternate policies with different level of co-optimization. 
    \begin{enumerate}
        \item Payment reduction policy (PR) : all devices are schedule under the rationale of minimizing the payment. Flexible loads are scheduled to maximize the surplus.
        \item Non co-optimization policy (NCO) : problem (11) is solved sequentially in the order of EV, consumption, and storage by limiting action space to the scheduling device and using the scheduled value.
        \item Charging-consumption co-optimized policy (CCO) : consumption and EV charging are
        co-optimized following Theorem 1, and storage is scheduled with the remaining DER considering the storage salvage value.
        \item MPC : solves the co-optimization problem (11) iteratively, with the updated state and predictions, and applies decisions for the first interval.
    \end{enumerate}

    \begin{figure*}[t]
         \centering
         \includegraphics[width = 0.8\textwidth]{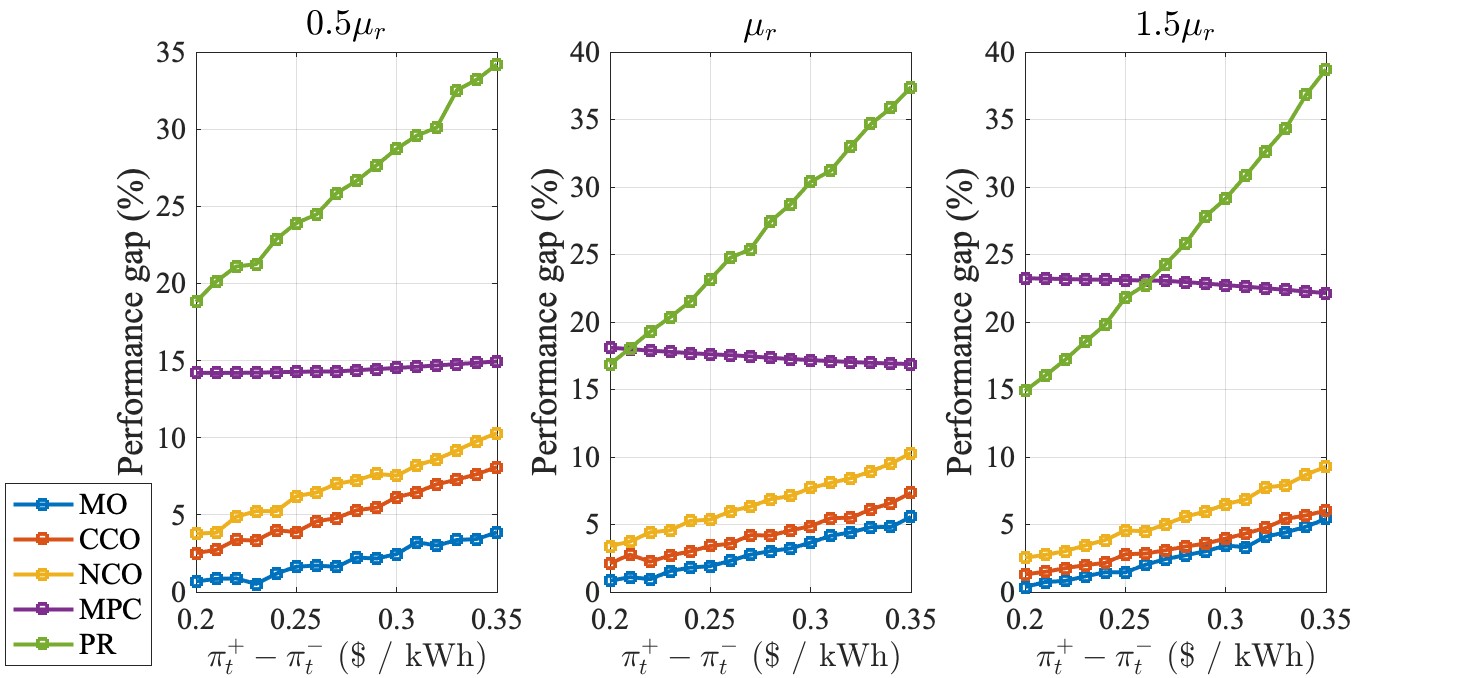}
         \caption{Relative performance gap of the MO(blue), CCO(red), NCO(yellow), MPC(purple), PR(green) to the oracle policy for different $\pi_t^+ - \pi_t^-$ is plotted. The estimated mean of renewable is scaled by 0.5, 1, 1.5 for the figures from left to right.}
         \label{fig7}
         \vspace{-1em}
     \end{figure*}
     
    \subsection{Simulation setting}
     We implemented Monte Carlo (MC) simulation with random EV charging demand, BTM DER trajectory, and connection time. The distribution of EV charging demand was modeled using charging demand data from Adaptive Charging Network \cite{lee_acndata_2019}. To model the distribution of rooftop solar generation, we used New York residential solar generation data from the Pecan Street \cite{pecanstreetdata}. Lastly, we assumed that the connection time of the EV is equally distributed.

    For the NEM ToU tariff parameters, we used the ToU retail rate from the Pacific Gas and Electronic (PG\&E)\footnote{PG\&E ToU tariff data can be found at \href{https://www.pge.com/tariffs/electric.shtml}{PGE E ToU-B}.} with on-peak hours from 4 PM to 9 PM. The sell rate was selected as a parameter to vary. The BTM storage had a maximum capacity of $B = 13.5$ kWh and maximum charging/discharging rates $\bar e = \underline e = 3.2$kW with charging/discharging inefficiency constants $\eta_c = \eta_d = 0.95$. We assumed that $s_0 = 0.5B$. For simplicity, we only considered a single controllable load, modeling total consumption of the household. We adopted a quadratic concave utility function that has the form $U(d) = a d - 1/2 b d^2$ which is independent of time. The coefficients of the utility function were estimated based on the consumption data from Pecan Street. The maximum charging capacity of EV charger is 3.6 kW, typical output power of level-2 charger.

    To obtain the expected accumulated reward for each policy, we ran 100,000 MC runs except for the MPC due to the computational limit. For the MPC, we ran 500 MC runs.

    \subsection{Performance of myopic optimal policy} 
    We present the performance gap of different policies when the retail rate is fixed and the sell rate is varied, assuming that $\pi_t^+ - \pi_t^-$ is kept constant at all intervals. The decision horizon $T$ is fixed as 16 hours. The three different plots in Fig.\ref{fig7} show the relative surplus gap when the mean of the distribution of DER generation is scaled by 0.5, 1, 1.5 from left to right. 
    
    In the plots, the MO has the smallest gap with the oracle policy ranging from 0.5-5\% in all scenarios. The performance gap increases as the mean of the renewable increases, because it is more likely that the SoC reaches the boundary as more energy is stored in the battery. We can also observe that the gap with MO to other control policies are shrinking as the mean  increases. When the renewable level is high, decisions made by non co-optimization polices and co-optimization policies will be indifferent because there will be sufficient renewables to allocate to every device.

\section{Conclusion}
This paper considers the co-optimization of BTM DER, storage, EV charging and controllable loads under NEM tariff. The main contribution is a myopic policy, which is optimal when the BTM storage schedules doesn't reach its boundary. The myopic policy is in general not optimal. It has been observed, however, that even when the storage schedule reaches the SoC limit, the myopic decision may still be optimal.  Our future work includes developing data-driven machine learning approaches that eliminate some of the model assumptions.
\vspace{-0.3em}

\bibliography{ref.bib}

\begin{thebibliography}{10}
\providecommand{\url}[1]{#1}
\csname url@samestyle\endcsname
\providecommand{\newblock}{\relax}
\providecommand{\bibinfo}[2]{#2}
\providecommand{\BIBentrySTDinterwordspacing}{\spaceskip=0pt\relax}
\providecommand{\BIBentryALTinterwordstretchfactor}{4}
\providecommand{\BIBentryALTinterwordspacing}{\spaceskip=\fontdimen2\font plus
\BIBentryALTinterwordstretchfactor\fontdimen3\font minus
  \fontdimen4\font\relax}
\providecommand{\BIBforeignlanguage}[2]{{%
\expandafter\ifx\csname l@#1\endcsname\relax
\typeout{** WARNING: IEEEtran.bst: No hyphenation pattern has been}%
\typeout{** loaded for the language `#1'. Using the pattern for}%
\typeout{** the default language instead.}%
\else
\language=\csname l@#1\endcsname
\fi
#2}}
\providecommand{\BIBdecl}{\relax}
\BIBdecl

\bibitem{solarstoragereport}
G.~Barbose, S.~Elmallah, and W.~Gorman, ``Behind-the-meter solar + storage :
  Market data and trends,'' Lawrence Berkeley National Laboratory, July 2021.

\bibitem{alahmed2022net}
A.~S. Alahmed and L.~Tong, ``On net energy metering {X}: Optimal prosumer
  decisions, social welfare, and cross-subsidies,'' \emph{IEEE Transactions on
  Smart Grid}, 2022.

\bibitem{alahmed2022co}
A.~S. Alahmed, L.~Tong, and Q.~Zhao, ``Co-optimizing distributed energy
  resources in linear complexity under net energy metering,'' \emph{arXiv
  preprint arXiv:2208.09781}, 2022.

\bibitem{jeon2022co}
M.~Jeon, L.~Tong, and Q.~Zhao, ``Co-optimizing consumption and {EV} charging
  under net energy metering,'' \emph{arXiv preprint arXiv:2211.11088}, 2022.

\bibitem{hafiz2019coordinated}
F.~Hafiz, A.~R. de~Queiroz, and I.~Husain, ``Coordinated control of {PEV} and
  {PV}-based storages in residential systems under generation and load
  uncertainties,'' \emph{IEEE Transactions on Industry Applications}, vol.~55,
  no.~6, pp. 5524--5532, 2019.

\bibitem{Seal2023}
S.~Seal, B.~Boulet, V.~R. Dehkordi, F.~Bouffard, and G.~Joos, ``Centralized
  {MPC} for home energy management with {EV} as mobile energy storage unit,''
  \emph{IEEE Transactions on Sustainable Energy}, pp. 1--10, 1 2023.

\bibitem{yousefi2020predictive}
M.~Yousefi, A.~Hajizadeh, M.~N. Soltani, and B.~Hredzak, ``Predictive home
  energy management system with photovoltaic array, heat pump, and plug-in
  electric vehicle,'' \emph{IEEE Transactions on Industrial Informatics},
  vol.~17, no.~1, pp. 430--440, 2020.

\bibitem{chen2013mpc}
C.~Chen, J.~Wang, Y.~Heo, and S.~Kishore, ``{MPC}-based appliance scheduling
  for residential building energy management controller,'' \emph{IEEE
  Transactions on Smart Grid}, vol.~4, no.~3, pp. 1401--1410, 2013.

\bibitem{xu2020multi}
X.~Xu, Y.~Jia, Y.~Xu, Z.~Xu, S.~Chai, and C.~S. Lai, ``A multi-agent
  reinforcement learning-based data-driven method for home energy management,''
  \emph{IEEE Transactions on Smart Grid}, vol.~11, no.~4, pp. 3201--3211, 2020.

\bibitem{Pedrasa2010}
M.~A.~A. Pedrasa, T.~D. Spooner, and I.~F. MacGill, ``Coordinated scheduling of
  residential distributed energy resources to optimize smart home energy
  services,'' \emph{IEEE Transactions on Smart Grid}, vol.~1, pp. 134--143, 9
  2010.

\bibitem{Erdinc2015}
O.~Erdinc, N.~G. Paterakis, T.~D. Mendes, A.~G. Bakirtzis, and J.~P. Catalão,
  ``Smart household operation considering bi-directional {EV} and {ESS}
  utilization by real-time pricing-based {DR},'' \emph{IEEE Transactions on
  Smart Grid}, vol.~6, pp. 1281--1291, 5 2015.

\bibitem{Hubert2012}
T.~Hubert and S.~Grijalva, ``Modeling for residential electricity optimization
  in dynamic pricing environments,'' \emph{IEEE Transactions on Smart Grid},
  vol.~3, pp. 2224--2231, 2012.

\bibitem{Li2018}
T.~Li and M.~Dong, ``Real-time residential-side joint energy storage management
  and load scheduling with renewable integration,'' \emph{IEEE Transactions on
  Smart Grid}, vol.~9, pp. 283--298, 1 2018.

\bibitem{Xu2017}
Y.~Xu and L.~Tong, ``Optimal operation and economic value of energy storage at
  consumer locations,'' \emph{IEEE Transactions on Automatic Control}, vol.~62,
  pp. 792--807, 2 2017.

\bibitem{Yu2020}
L.~Yu, W.~Xie, D.~Xie, Y.~Zou, D.~Zhang, Z.~Sun, L.~Zhang, Y.~Zhang, and
  T.~Jiang, ``Deep reinforcement learning for smart home energy management,''
  \emph{IEEE Internet of Things Journal}, vol.~7, pp. 2751--2762, 4 2020.

\bibitem{Kim2011}
T.~T. Kim and H.~V. Poor, ``Scheduling power consumption with price
  uncertainty,'' \emph{IEEE Transactions on Smart Grid}, vol.~2, pp. 519--527,
  9 2011.

\bibitem{wu2018optimizing}
D.~Wu, G.~Rabusseau, V.~Fran{\c{c}}ois-lavet, D.~Precup, and B.~Boulet,
  ``Optimizing home energy management and electric vehicle charging with
  reinforcement learning,'' \emph{Proceedings of the 16th Adaptive Learning
  Agents}, 2018.

\bibitem{jeon2023Procras}
M.~Jeon, L.~Tong, and Q.~Zhao, ``On the optimality of procrastination policy
  for ev charging under net energy metering,'' \emph{arXiv preprint
  arXiv:2304.04076}, 2023.

\bibitem{lee_acndata_2019}
Z.~J. Lee, T.~Li, and S.~H. Low, ``{ACN}-{Data}: {Analysis} and {Applications}
  of an {Open} {EV} {Charging} {Dataset},'' in \emph{Proceedings of the Tenth
  International Conference on Future Energy Systems}, ser. e-Energy '19, Jun.
  2019.

\bibitem{pecanstreetdata}
``Pecan street dataset,'' Available at \url{www.pecanstreet.org/dataport/}
  (2022/11/01).

\end{thebibliography}
\bibliographystyle{IEEEtran}

\clearpage
\appendices
\section{Proofs}
The optimality of procrastination threshold policy for the co-optimization without storage (Theorem 1) was proved \cite{jeon2022co}. We will prove the Proposition 1 which wasn't stated in \cite{jeon2022co}.

\subsection{Proof of \normalfont{Proposition 1}}
    \begin{proof}
    By the proposition 2 in \cite{jeon2022co}, $\tau_t $ and $\tau_{t+1}$ satisfy : 
    \begin{equation*}
     \tau_t = \tau_{t+1} + \bar v   
    \end{equation*}
    For some $t_0 \in \mathcal T_x$, given $y_{t_0}$, if $z_{t_0}^* v_{t_0}^* > 0$, by the procrastination threshold policy, $y_{t_0} > \tau_{t_0}$ and,
    \begin{equation*}
        v_{t_0}^* = \min\{\bar v, y_{t_0} - \tau_{t_0}\}.    
    \end{equation*}
    Then, $y_{t_0+1} = y_{t_0} - v_{t_0}^* \ge \tau_{t_0} = \tau_{t_0+1} + \bar v.$ Hence, by the monotonicity of $v_t^*$ with respect to $r_t$, $v_{t_0+1}^* = \bar v$ for all $r_{t_0+1}$. 
    
    The recursive relation of $\tau_t$ holds within the same period, therefore, $v_t^* = \bar v$ for all $t \in \mathcal T_0$ such that $t > t_0$. 
    \end{proof}

\subsection{Concavity of the optimal value function}
    We first prove the concavity of the optimal value function of myopic co-optimization. Here, we are only interested in concavity of the value function with respect to $y_t$. So, let's simplify the notation as $V_t^M(y_t)$.
    \begin{lemma}
        Optimal value function of myopic co-optimization, $V_t^M(y_t)$, that satisfies Bellman equation (\ref{eq:myopic_opt}) is concave function of $y_t$.
    \end{lemma}
\begin{proof}
    At $t = T-1$, there is no uncertainty and Bellman equation becomes
    \begin{equation*}
        \begin{aligned}
            V_{T-1}^M(y_{T-1}) &= \max_{v,e, \mathbf d} \, \{ g_{T-1}(x_{T-1}, v, e, \mathbf d) - \alpha (y_{T-1} - v) \\
            & \quad + \beta (\mathbbm 1_{e \ge 0} \eta_c - \mathbbm 1_{e \le 0}/\eta_d) (s_{T-1} + e) \} \\
            &\text{s.t.} \quad (4)-(5)
        \end{aligned}
    \end{equation*}
    As $\alpha > \pi^+ > \pi^-$, the optimal EV charging action is 
    \begin{equation*}
        v^* = \min\{ \bar v, y_{T-1}\}.
    \end{equation*}
    For $y_{T-1} > \bar v$, optimal consumption and storage operation are independent to $y_{T-1}$ and $\frac{\partial}{\partial y_{T-1}}V^M_{T-1} = - \alpha$. 

    For $y_{T-1} \le \bar v$, optimal consumtion and storage operation is decided based on the remaining renewables $\tilde r_{T-1} = \max \{ r_{T-1} - y_{T-1}, 0\}$. Consumption-storage co-optimization decision is based on myopic co-optimization in \cite{alahmed2022co}. 
    
    If household is net consuming, $\tilde r_{T-1} \le l_{T-1}(\pi_{T-1}^+) - \underline e' $, $\frac{\partial}{\partial y_{T-1}}V^M_{T-1} = - \pi_{T-1}^+ < - \alpha$.

    As $y_{T-1}$ reduces, $\tilde r_{T-1}$ increases, and household will be in the net zero zone. In the net zero zone, $\frac{\partial}{\partial y_{T-1}}V^M_{T-1} $ is the negative of marginal utility of consumption, and as the consumption is increasing, marginal utility decreases. Therefore, $\frac{\partial}{\partial y_{T-1}}V^M_{T-1} $ increases. At last, household enters, net producing region, which means $\frac{\partial}{\partial y_{T-1}}V^M_{T-1} = -\pi_{T-1}^-$. 

    Hence, $\frac{\partial}{\partial y_{T-1}}V^M_{T-1} $ is a decreasing function of $y_{T-1}$, therefore, $V^M_{T-1}(x_{T-1})$ is a concave function of $y_{T-1}$.

    Suppose, $V^M_{t+1}$ is a concave function of $y_{t+1}$. Then, the objective function of bellman equation is concave and continuous in the compact feasible set, there exists a optimal solution for every $y_t$. Let's denote the optimal solution for $y_t = \kappa, $ as $v_\kappa^*, e_\kappa^*, \mathbf d_\kappa^*$, and for $y_t = \zeta$ as $v_\zeta^*, e_\zeta ^*, \mathbf d_\zeta ^*$. 
    \begin{align*}
        \lambda V_t^M(\kappa) &+ (1 - \lambda) V_t^M(\zeta) \\
        &= \lambda \{g_t(\kappa, v_\kappa^*, e_\kappa^*, \mathbf d_\kappa^*) + \mathbb E[V_{t+1}^M(\kappa - v_\kappa^*)]\} \\
        &+(1 - \lambda) \{g_t(\zeta, v_\zeta^*, e_\zeta ^*, \mathbf d_\zeta ^*) + \mathbb E[V_{t+1}^M(\zeta - v_\zeta^*)\} \\
        &\le g_t(\xi, v_\xi, e_\xi, \mathbf d_\xi) + \mathbb E[V_{t+1}^M(\xi - v_\xi)] \\
        &\le g_t(\xi, v_\xi^*, e_\xi^*, \mathbf d_\xi^*) + \mathbb E[V_{t+1}^M(\xi - v_\xi^*)] \\
        &= V_t^M(\xi) = V_t^M(\lambda \kappa + (1 - \lambda) \zeta)
    \end{align*}
    Here, $\xi = \lambda \kappa + (1 - \lambda) \zeta$, $v_\xi = \lambda v_\kappa^* + (1 - \lambda)  v_\xi^*$, $e_\xi = \lambda e_\kappa^* + (1 - \lambda)  e_\xi^*$, $\mathbf d_\xi = \lambda \mathbf d_\kappa^* + (1 - \lambda)  \mathbf d_\xi^*$. \\
    
    The first inequality holds, because of the concavity of the stage reward function of inductive hypothesis. 

    Hence, $V_t^M$ is a concave function of $y_t$ for all $t \in \mathcal T$.
\end{proof}

 \subsection{Storage operation propositions}
 For the rest of the proofs, $d_t^* = \mathbf 1^T\mathbf d_t^*$. The proof of the following propositions are shown in \cite{alahmed2022co}
\begin{proposition}[Storage salvage value under non-binding SoC assumption]
    Under assumption A1 and A2, for $\epsilon > 0$ that doesn't violate non-binding SoC assumption, 
    \begin{align*}
        V_t(s_t + \epsilon, y_t, r_t) &= V_t(s_t, y_t, r_t) + \beta \eta_c \epsilon \\
        V_t(s_t -\epsilon, y_t, r_t) &= V_t(s_t, y_t, r_t) - \beta / \eta_d\epsilon
    \end{align*}
\end{proposition}

\begin{proposition}[Storage-total load complementarity condition]
    Under A1-A2, optimal storage operations and net consumption of the myopic co-optimization (\ref{eq:myopic_opt}) satisfy 
    \begin{gather*}
        e_t^*z_t^* \le 0,\\
        e_t^* (d_t^* + v_t^* - r_t)\le 0.
    \end{gather*}
\end{proposition}

\begin{proposition}[Optimal storage operation]
    Under A1-A2, for all $t\in \mathcal T$, for the optimal consumption $\mathbf d_t^*$ and EV charging $v_t^*$ of (\ref{eq:myopic_opt}), $e_t^*$ satisfies 
    \begin{equation*}
        e^*_t = \begin{cases}
            \max\{r_t - d_t^* - v_t, -\underline e'\}, & r_t \le d_t^* + v_t^* \\
            \min \{ r_t - d_t^* - v_t, \bar e'\}, & r_t > d_t^* + v_t^*
        \end{cases}
    \end{equation*}
    and $e_t^* \le r_t$.
\end{proposition}

\subsection{Proof of \normalfont{Theorem 2}}
\begin{proof}
Let's consider three regions of $r_t$ divided by $\Delta_t^{+'}$ and $\Delta_t^{-'}$.
\begin{enumerate}
    \item $r_t < \Delta_t^{+'}(y_t)$ : Define storage output augmented renewable $\tilde r_t := r_t - e_t^* $. Then, by Proposition 5, and $e_t^* \ge -\underline e'$, 
    \begin{equation*}
        0 < \tilde r_t < l_t(\pi_t^+) + h_{\tau_t}(y_t)    
    \end{equation*}
    Hence, by Theorem 1, 
    \begin{equation*}
        v_t^* = h_{\tau_t}(y_t), \, d_{ti}^* = l_{ti}(\pi_t^+), \; \forall i = 1, \ldots, K.
    \end{equation*}
    Then, $v_t^* + d_{t}^* > r_t$, and $e_t^* = -\underline e'$. Therefore, 
    \begin{equation*}
        z_t^* = v_t^* + d_t^* + e_t^* - r_t = \Delta_t^{+'}(y_t) - r_t
    \end{equation*}
    \item $r_t > \Delta_t^{-'}(y_t)$ : For such $r_t$, storage output augmented renewable satisfies 
    \begin{equation*}
        \tilde r_t  = r_t - e_t^* > l_t(\pi_t^-) + h_{\delta_t}(y_t)
    \end{equation*}
    Then, by Theorem 1, 
    \begin{equation*}
        v_t^* = h_{\delta_t}(y_t), \, d_{ti}^* = l_{ti}(\pi_t^-), \; \forall i = 1, \ldots, K.
    \end{equation*}
    From Proposition 5, as $v_t^* + d_t^* < r_t$, $e_t^* = \bar e'$. Therefore, 
    \begin{equation*}
       z_t^* = v_t^* + d_t^* + e_t^* - r_t = \Delta_t^{-'}(y_t) - r_t
    \end{equation*}
    \item $r_t \in [\Delta_t^{+'}(y_t), \Delta_t^{-'}(y_t)]$ : Note that $z_t^* = 0$ for $r_t = \Delta_t^{+'}(y_t)$ and $r_t = \Delta_t^{-'}(y_t)$. Hence, it's suffice to show that $z_t^*$ is a monotone decreasing function of $r_t$ in this region.\\
    
    For $r_t = \Delta_t^{+'}(y_t)$, and $\epsilon > 0$ such that $r_t + \epsilon \in[\Delta_t^{+'}(y_t), \Delta_t^{-'}(y_t)] $, suppose optimal net consumption of $r_t + \epsilon$ is $\tilde z_t^* > 0$. Let's denote the optimal scheduling of each device as $\tilde v_t^*, \, \tilde {\mathbf d}_t^*, \, \tilde e_t^*$. By Proposition 5, $\tilde e_t^* = -\underline e'$. 
    \begin{equation*}
        r_t + \epsilon - \tilde e_t^* = r_t + \epsilon + \underline e' \ge r_t - e_t^* = l_t(\pi_t^+) + h_{\delta_t}(y_t) = \Delta_t^+(y_t)
    \end{equation*}
    The first inequality holds because $e_t^* \ge -\underline e'$. The second equality comes from Theorem 1.\\
    
    By Theorem 1, if $r_t + \epsilon - \tilde e_t^* \ge \Delta_t^+(y_t)$, $\tilde v_t^* + \tilde{\mathbf d}_t^* \le r_t +\epsilon - \tilde e_t^*$. Hence, $\tilde z_t^* \le 0$ which contradicts the assumption. Therefore, $z_t^*$ is a monotone decreasing function with respect to $r_t$.
\end{enumerate}

    Now let's show that $v_t^*, d_t^*, e_t^*$ are determined by \textbf{Algorithm} \ref{alg:Myopic_optimal_algorithm}. From the proof of optimal net consumption, we proved optimal myopic decision for $r_t < \Delta_t^{+'}(y_t)$ and $r+t > \Delta_t^{-'}(y_t)$. Let's prove the myopic optimal schedule in the net-zero zone. In this region, myopic co-optimization becomes, 
    \begin{equation*}
    \begin{aligned}
    &\max_{v,e,\mathbf d}  && U_t(\mathbf d) +\big(\eta_c \mathbbm 1_{e_t \ge 0} - \mathbbm 1_{e_t\le 0} / \eta_d\big)\beta e_t &\\
    & & & +  \mathbbm E[V^M_{t+1}(s_t,y_t - v, r_{t+1}, \pi_{t+1})]  &\\
    &\text{s.t.} &&  -\underline e' \le e_t \le \bar e' &(\bar \lambda_e, \underline \lambda_e)\\
    & && 0 \preceq \mathbf d \preceq \bar{\mathbf d} &( \boldsymbol{\bar\lambda}_d, \boldsymbol{\underline \lambda}_d)\\
    & &&  0 \le v \le \bar v &(\bar \lambda_v, \underline \lambda_v)\\
    & && v + \mathbf 1 ^T \mathbf d + e = r & (\nu)
    \end{aligned}
    \end{equation*}
    Here, $(\bar \lambda_e, \underline \lambda_e), \, ( \boldsymbol{\bar\lambda}_d, \boldsymbol{\underline \lambda}_d), \,(\bar \lambda_v, \underline \lambda_v)$ are the Lagrange multipliers for the corresponding constraints. 
    
    Lagrangian of the optimization problem is 
    \begin{align*}
        \mathcal L &= U(\mathbf d) + \mathbbm E[V^M_{t+1}(s_t,y_t - v, r_{t+1}, \pi_{t+1})]  \\
        &+\big(\eta_c \mathbbm 1_{e_t \ge 0} - \mathbbm 1_{e_t\le 0} / \eta_d\big)\beta e_t + \underline \lambda_e(e_t + \underline e') + \bar \lambda_e(\bar e' - e_t) \\
        &+ \boldsymbol{\underline \lambda}_d^T\mathbf d + \boldsymbol{\bar \lambda }_d^T(\mathbf{\bar d} - \mathbf d) + \underline \lambda _v v + \bar \lambda_v (\bar v - v) \\
        &+ \nu(v + \mathbf 1 ^T \mathbf d + e - r)
    \end{align*}
    As above optimization satisfies Slater's condition, KKT condition becomes necessary and sufficient condition. Let's verify myopic optimal decisions in \textbf{Algorithm}~\ref{alg:Myopic_optimal_algorithm} satisfies KKT condition.
    \begin{enumerate}
        \item $r_t \in \big[\Delta_t^{+'}(y_t), \Delta_{t,1}(y_t)\big) : $ By complementary slackness, $\underline \lambda_e = 0$. Then, $\nu = \beta / \eta_d + \bar \lambda_e$.\\
        
        $\tilde r_t = r_t + \underline e' \in [\Delta_t^+(y_t), l_t(\beta / \eta_d) + h_{\sigma_t^+}(y_t))$. Hence, from Theorem 1, optimal EV charging and consumption decisions are
        \begin{equation*}
            v_t^* = h_{w_{t+1}(\nu)}(y_t), \, d_{ti}^* = l_{ti}(\nu), \, \forall i = 1, \ldots, K
        \end{equation*}
        where $\nu \in [\pi_t^+, \beta / \eta_d)$. Here, due to the monotonicity of $L_t$ and $\partial (\bar V_t)$, $v_t^*$ and $d_t^*$ are monotone increasing with respect to $r_t$.
        
        \item $r_t \in \big[\Delta_{t,1}(y_t), \Delta_{t,2}(y_t) \big)$ : By complementary slackness, $\bar \lambda_e = \underline \lambda_e = 0$. Hence, $\nu = \beta / \eta_d$. Then, by the stationarity condition, optimal EV charging and consumption decisions are 
        \begin{equation*}
            v_t^* = h_{\sigma_t^+}(y_t), \, d_{ti}^* = l_{ti}(\beta / \eta_d), \, \forall i = 1, \ldots, K,
        \end{equation*}
        where $\sigma_t^+ = w_{t+1}(\beta / \eta_d)$. By monotonicity of $L_t$ and $\partial (\bar V_t)$, optimal EV charging and consumption decisions are no less than the optimal EV charging and consumption decisions of previous case, respectively. Hence, montonicity of myopic optimal policy still holds. 
        
        \item $r_t \in \big[\Delta_{t,2}(y_t), \Delta_{t,3}(y_t) \big)$ : As $e_t^* = 0$, by complementary slackness, $\bar \lambda_e = \underline \lambda_e = 0$. From Theorem 1, optimal EV charging and consumption decisions are 
        \begin{equation*}
            v_t^* = h_{w_{t+1}}(y_t), \, d_{ti}^* = l_{ti}(\nu), \; \forall i = 1, \ldots, K
        \end{equation*}
        where $\nu \in [\beta \eta_c, \beta / \eta_d]$ and $v_t^* + d_t^* = r_t$. Similarly, monotonicity of $v_t^*$ and $d_t^*$ holds due to monotonicity of $L_{ti}$ and $\partial (\bar V_t)$.
        
        \item  $r_t \in \big[\Delta_{t,3}(y_t), \Delta_{t,4}(y_t) \big)$ : By complementary slackness, $\bar \lambda_e = \underline \lambda_e = 0$ and $\nu = \eta_c \beta$. By stationarity condition, optimal EV charging and consumption decisions are 
        \begin{equation*}
             v_t^* = h_{\sigma_t^-}(y_t), \, d_{ti}^* = l_{ti}(\beta / \eta_d), \, \forall i = 1, \ldots, K,
        \end{equation*}
        where $\sigma_t^- = w_{t+1}(\beta \eta_c)$. Montonicity of the optimal decisions still holds.
        
        \item $r_t \in \big[\Delta_{t,4}(y_t), \Delta_{t}(y_t)^{-'} \big)$ : By complemenatry slackness, $\bar \lambda_e = 0$. Then, $\nu = \beta \eta_c + \underline \lambda_e$. 
        
        For $\tilde r_t = r_t - \bar e' \in \big [l_t(\beta \eta_c) + h_{\sigma_t^-}(y_t), \Delta_t^-(y_t)\big)$, by Theorem 1, optimal EV charging and consumption decisions are 
        \begin{equation*}
            v_t^* = h_{w_{t+1}}(y_t), \, d_{ti}^* = l_{ti}(\nu), \; \forall i = 1, \ldots, K,
        \end{equation*}
        where $\nu \in [\beta \eta_c, \pi_t^-]$. 
    \end{enumerate}
    Overall, monotonicity of $v_t^*, \, d_t^*$ holds due to the monotonicity of $l_{ti}$ and $\partial(\bar V_t)$. 
\end{proof}

\subsection{Proof of Theorem 3}
\begin{proof}
It's suffice to show that for $s_t$ that satisfies A2 and $\boldsymbol \gamma = (\epsilon, 0, 0, 0) $ ($\epsilon > 0$), optimal value function that satisfies (\ref{eq:dp}), satisfies 
\begin{align}\label{eq:myopic_condition}
    V_t(x_t + \boldsymbol \gamma) &= V_t(x_t) + \beta \epsilon
\end{align}
If $V_t$ satisfies above equations, $V_t$ reduces to $V_t^M$. Let's show above equations by backward induction. For $t = T-1$, 
\begin{gather*}
    V_{T-1}(x_{T-1} + \boldsymbol \gamma) = \max_{v,e, \mathbf d} \; \{g_{T-1}(x_{T-1} + \boldsymbol \gamma , v, e, \mathbf d) \\
    + \beta (s_{T-1} + \epsilon + \eta_c\mathbbm 1_{e_t \ge 0}e - \mathbbm 1_{e_t < 0} / \eta_d)e) + \alpha (y_{T-1} - \alpha v) \} \\
    = \max_{v,e ,\mathbf d} \; \{g_{T-1}(x_{T-1}, v, e, \mathbf d) + \beta (s_{T-1}+ \eta_c\mathbbm 1_{e_t \ge 0}e - \mathbbm 1_{e_t < 0} / \eta_d)e) \\
    + \alpha (y_{T-1} - \alpha v) \}  + \beta \epsilon = V_{T-1}(x_{T-1}) + \beta\epsilon.
\end{gather*}
Suppose $V_{t+1}$ satisfies (\ref{eq:myopic_condition}), then, from (\ref{eq:dp}),
\begin{align*}
    V_t(x_t + \boldsymbol \gamma)  &= \max_{v,e, \mathbf d} \; \{g_{t}(x_{t} + \boldsymbol \gamma , v, e, \mathbf d) + \mathbbm E[ V_{t+1}(x_{t+1} + \boldsymbol \gamma )]\} \\
    &= \max_{v,e, \mathbf d} \{g_{t}(x_{t}, v, e, \mathbf d)  + \mathbbm E[ V_{t+1}(x_{t+1})] \}+ \beta \epsilon\\
    &= V_t(x_t) + \beta \epsilon.
\end{align*}
\end{proof}

\subsection{Proof of Proposition 2}
\begin{proof}
Procrastination thresholds $\tau_t$ and $\delta_t$ follows from the Theorem 2 in \cite{jeon2022co}. Also, characterization of $\sigma_t^+$ and $\sigma_t^-$ comes from the proof of Theorem 2. 

Also, recursive relation of $\tau_t$ and $\delta_t$ comes from the proof of Proposition 1 in \cite{jeon2022co}.
\end{proof}

\vfill\null
%\columnbreak
\section{Algorithms}

Algorithm 1 presented in Theorem 2 that decides myopic optimal scheduling decisions is presented here. 
    \begin{algorithm} [H]
    \caption{Myopic optimal decisions}
    \label{alg:Myopic_optimal_algorithm} 
    \begin{algorithmic}
    \Require $y_t, \pi_t, r_t$
    \Ensure $v_t^*, e_t^*, d_t^*$
    \State For all $i = 1, \ldots, K$, myopic optimal decisions of each device are : \\
    (with a slight abuse of notation, $\bar V_t(y) := \mathbb E[V_t^M(y,r_t,\pi_t))]$, \\
    $w_t(\pi) := (\partial \bar V_t)^{-1}(-\pi)$)
 %   \hspace*{\algorithmicindent}$w_t(\pi) := (\partial \bar V_t)^{-1}(-\pi)$) : 
        \If {$r_t < \Delta_t^{+'}(y_t)$}
            \State {$v_t^* \gets h_{\tau_t}(y_t), \quad
            d_{ti}^*  \gets  l_{ti}(\pi_t^+),  \quad
            e_t^* \gets -\underline  e'$},
        \ElsIf {$r_t \in \big[\Delta_t^{+'}(y_t), \Delta_{t,1}(y_t) \big)$}
            \State $v_t^* \gets h_{w_{t+1}(\nu)}(y_t), \quad 
            d_{ti}^* \gets l_{ti}(\nu), \quad 
            e_t^* \gets -\underline e'$, \newline
             where $r_t + \underline e' = v_t^* + \sum_{i=1}^Kd_{ti}^*$, and $\nu \in [\beta / \eta_d, \pi_t^+]$,
        \ElsIf {$r_t \in \big[\Delta_{t,1}(y_t), \Delta_{t,2}(y_t) \big)$}
            \State $v_t^* \gets h_{\sigma_t^+}(y_t), \quad
            d_{ti}^* \gets l_{ti}(\beta /\eta_d), \quad 
            e_t^* \gets r_t - \Delta_{t,2}$,
        \ElsIf{$r_t \in \big[\Delta_{t,2}(y_t), \Delta_{t,3}(y_t) \big)$}
          \State $v_t^* \gets h_{w_{t+1}(\nu)}(y_t), \quad 
            d_{ti}^* \gets l_{ti}(\nu), \quad 
            e_t^* \gets 0$, \newline
            where $r_t  = v_t^* + \sum_{i=1}^K d_{ti}^*$,
            and $\nu \in [\beta  \eta_c, \beta / \eta_d]$
        \ElsIf{$r_t \in \big[\Delta_{t,3}(y_t), \Delta_{t,4}(y_t) \big)$}
            \State $v_t^* \gets h_{\sigma_t^-}(y_t), \quad
            d_{ti}^* \gets l_{ti}(\beta\eta_c), \quad 
            e_t^* \gets r_t - \Delta_{t,3}$,
        \ElsIf{$r_t \in \big[\Delta_{t,4}(y_t), \Delta_{t}^{-'}(y_t) \big)$}
            \State $v_t^* \gets h_{w_{t+1}(\nu)}(y_t), \quad 
            d_{ti}^* \gets l_{ti}(\nu), \quad 
            e_t^* \gets \bar e'$, \newline
            where $r_t - \bar e'  = v_t^* + \sum_{i=1}^K d_{ti}^*$,
            and $\nu \in [\pi_t^-,\beta  \eta_c]$
        \Else
            \State $v_t^* \gets h_{\delta_t}(y_t), \quad 
            d_{ti}^* \gets l_{ti}(\pi_t^-), \quad
            e_t^* \gets \bar e'$ 
        \EndIf
    \end{algorithmic}
    \end{algorithm}

\end{document}